\documentclass{pasj00}
\draft
\usepackage{longtable}

\newcommand{\cjaa}{Chinese J. Astron. Astrophys.}

\newcommand{\na}{NewA.  }

\begin{document}
\SetRunningHead{Author(s) in page-head}{Running Head}

\title{The W-subtype active contact binary PZ UMa with a possible more massive tertiary component}



%
 \author{\textsc{X.} Zhou\altaffilmark{1,2,3,4},
         \textsc{B.} Soonthornthum\altaffilmark{2}}
\altaffiltext{1}{Yunnan Observatories, Chinese Academy of Sciences (CAS), P.O. Box 110, 650216 Kunming, P. R. China}
 \email{zhouxiaophy@ynao.ac.cn}
\altaffiltext{2}{National Astronomical Research Institute of Thailand, 260  Moo 4, T. Donkaew,  A. Maerim, Chiangmai, 50180, Thailand}
\altaffiltext{3}{Key Laboratory of the Structure and Evolution of Celestial Objects, Chinese Academy of Sciences, P. O. Box 110, 650216 Kunming, P. R. China}
\altaffiltext{4}{Center for Astronomical Mega-Science, Chinese Academy of Sciences, 20A Datun Road, Chaoyang Dis-trict, Beijing, 100012, P. R. China}

\maketitle

\begin{abstract}
Two sets of multiple-color ($B, V, R_c, I_c$) light curves of PZ UMa were observed in dependently with the 2.4 meter telescope at the Thai National Observatory and the 1 meter telescope at Yunnan Observatories. The light curves were analyzed with the Wilson-Devinney program and the two sets of light curves produced consistent results, which show that PZ UMa is a W-subtype contact binary with an extreme mass ratio ($M_{1}/M_{2} = 0.18)$. The basic physical parameters of PZ UMa were determined to be $M_{2} = 0.77(2)M_\odot$, $M_{1} = 0.14(1)M_\odot$, $R_{2} = 0.92(1)R_\odot$, $R_{1} = 0.43(1)R_\odot$, $L_{2} = 0.46(2)L_\odot$ and $L_{1} = 0.15(3)L_\odot$. The orbital period analysis of PZ UMa revealed a 13.22 year periodicity, which implies that there may be a tertiary component orbiting around the binary system. The mass and orbital radius of the tertiary component were calculated to be $M_{3} = 0.88 M_\odot$ and $a_{3} = 3.67 AU$, if the orbit was coplanar with the central binary system. It is interesting that the minimum mass of the tertiary was calculated to be $M_{3min} = 0.84 M_\odot$, which means the tertiary component is even larger than the primary star and the secondary one of PZ UMa. PZ UMa is a late-type contact binary with stellar activity. The O'Connell effect appeared on its light curves when it was observed on April 2016. However, the O'Connell effect reversed when the target was observed again on December 2016. The changes of the O'Connell effect in such a short time-scale strongly support the occurrence of rapidly changing magnetic activity on this W UMa binary.
\end{abstract}

\section{INTRODUCTION}

Photometric observations of eclipsing binaries have a long history. In 1955, \citet{1955AnAp...18..379K} divided close binaries into three categories, which were detached binaries, semi-detached binaries and contact binaries, depending on whether the component stars had filled their critical Roche lobes. They usually have EA, EB and EW type light curves correspondingly. As for contact binaries, both of the component stars have filled their critical Roche lobes. Contact binary systems share a common envelope and mass transfer occurs between the two components. This peculiar structure makes the evolution of contact binaries very complex and interesting. In 1970, \citet{1970VA.....12..217B} found that there were two subtypes of contact binaries, which were A-subtype and W-subtype. For the A-subtype system, the more massive component star has a higher surface temperature, while the less massive star is the hotter one in a W-subtype system. It is really striking to researchers that a less massive star can be the hotter one in a binary system. The formation theory of W-subtype contact binaries is still unclear. \citet{2013MNRAS.430.2029Y} pointed out that W-subtype contact binaries have initial masses larger than 1.8$M_\odot$ and subsequent research by \citet{2016ApJ...817..133Z} also supported their conclusions. More photometric/spectroscopic observations and analyses of W-subtype contact binaries are needed to further characterize them.

The light variations of PZ UMa (other names: GSC 03428-00212, NSVS 4861481) were firstly reported by \citet{2005OEJV...12....1N} while searching for new variable stars in the public data of the Northern Sky Variability Survey \citep{2004AJ....127.2436W}. It was identified as an EW type contact binary with its an orbital period of 0.2627 days. Photometric observations of PZ UMa were carried out using the optical telescopes at the Thai National Observatory and Yunnan Observatories. Two sets of multiple-color light curves were obtained to study its light variations and period changes. The effective temperature given in the Gaia Data Release 2 \citep{2016A&A...595A...1G,2018A&A...616A...1G} is $T = 5430(260)K$, which means PZ UMa is a G7 type contact binary \citep{Cox2000}.

\section{OBSERVATIONS AND DATA REDUCTION}

Observations of PZ UMa were carried out on 19 and 20 April 2016 with the 2.4 meter telescope at Thai National Observatory, National Astronomical Research Institute of Thailand (hereafter TNO 2.4m). The TNO 2.4m was equipped with a 4K $\times$ 4K CCD camera, and its field of view was 16 $\times$ 16 square arc-minutes \citep{2018NatAs...2..355S}. Johnson-Cousins filters were used during the observations.

PZ UMa was observed again on 10 December 2016 with the 1 meter telescope at Yunnan Observatories, Chinese Academy of Sciences (hereafter YNOs 1m). The YNOs 1m was equipped with a Andor DW436 2K $\times$ 2K CCD camera, and the field of view was 7.3 square arc-minutes. Like filters were used with the TNO 2.4m. On 29 November 2018 PZ UMa was observed with the 60 centimeter telescope at Yunnan Observatories (hereafter YNOs 60cm) in order to determine the mid eclipse time and the nature of the eclipses.

\begin{figure}[!ht]
\begin{center}
\includegraphics[width=8cm]{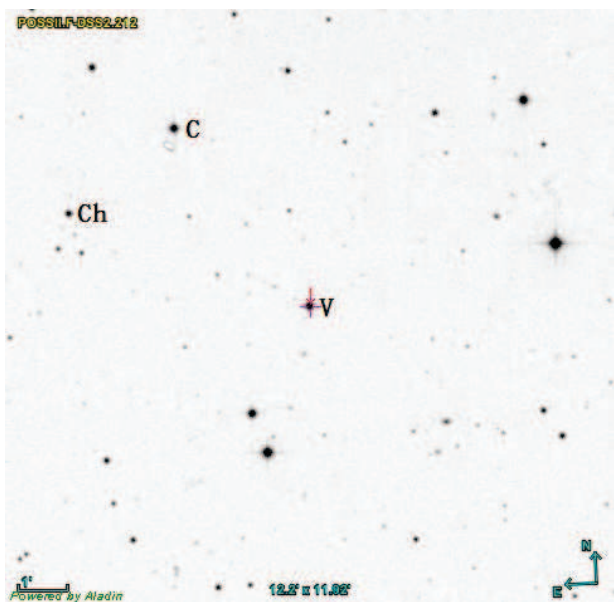}
\caption{Star map of PZ UMa with its field of view to be 12.2 $\times$ 11.92 arc-minutes.}\label{Image}
\end{center}
\end{figure}

\begin{figure}[!ht]
\begin{center}
\includegraphics[width=12cm]{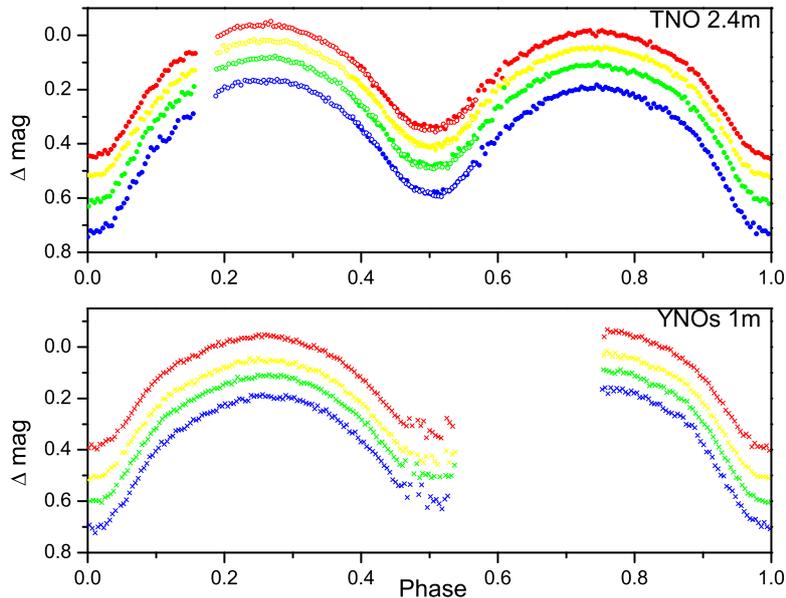}
\caption{Solid and open circles represent light curves of PZ UMa obtained with the TNO 2.4m on 19 and 20 April 2016, respectively. Crosses refer to light curves obtained with YNOs 1m on 10 December 2016. The blue, green, yellow and red colors denote light curves observed with $B$, $V$, $R_c$ and $I_c$ filters respectively.}\label{LC_Obs}
\end{center}
\end{figure}

The Image Reduction and Analysis Facility (IRAF) \citep{1986SPIE..627..733T} was used to reduce the observational images. UCAC4 700-051550 and UCAC4 700-051554 in the same field of view were selected as the Comparison (C) and Check star (Ch) respectively, in order to determine the light variations of PZ UMa. Their coordinates and $V$ band magnitudes are listed in Table \ref{Coordinates}. The image from the \textsl{Aladin Sky Atlas} \citep{2000A&AS..143...33B} that we used to identify the targets is given in Fig. \ref{Image}, in which PZ UMa (V), UCAC4 700-051550 (C) and UCAC4 700-051554 (Ch) are indicated. We performed parabola fits on the observed light minima in $B$, $V$, $R_C$ and $I_C$ bands and got the averaged mid-eclipse times. In all, two primary (p) and three secondary (s) minima were detected, and these are listed in Table \ref{New_minimum}. The phased light curves of PZ UMa are displayed in Fig. \ref{LC_Obs}.

\begin{table}[!h]\small
\begin{center}
\caption{Coordinates and $V$ band magnitudes.}\label{Coordinates}
\begin{tabular}{cccc}\hline\hline
Target              &   $\alpha_{2000}$         &  $\delta_{2000}$          &  $V_{mag}$   \\ \hline\hline
PZ UMa              &$09^{h}29^{m}07^{s}.078$   & $+49^\circ51'23''.065$    &  $12.705$     \\
UCAC4 700-051550    &$09^{h}29^{m}23^{s}.579$   & $+49^\circ54'59''.310$    &  $12.512$     \\
UCAC4 700-051554    &$09^{h}29^{m}36^{s}.855$   & $+49^\circ53'18''.86$     &  $14.060$     \\
\hline\hline
\end{tabular}
\end{center}
\end{table}

\begin{table}[!ht]\small
\begin{center}
\caption{New CCD times of mid-eclipse times.}\label{New_minimum}
\begin{tabular}{cccccc}\hline\hline
    JD (Hel.)     &  Error (days)  &  p/s &           Filter          &   Telescope\\\hline\hline
  2457498.0700    & $\pm0.0002$    &   s  &   $B$ $V$ $R_C$ $I_C$     &    TNO 2.4m   \\
  2457498.2008    & $\pm0.0001$    &   p  &   $B$ $V$ $R_C$ $I_C$     &    TNO 2.4m   \\
  2457499.1209    & $\pm0.0001$    &   s  &   $B$ $V$ $R_C$ $I_C$     &    TNO 2.4m   \\
  2457733.2925    & $\pm0.0001$    &   p  &   $B$ $V$ $R_C$ $I_C$     &    YNOs 1m   \\
  2458452.3552    & $\pm0.0002$    &   s  &           $R_C$ $I_C$     &    YNOs 60cm   \\
\hline\hline
\end{tabular}
\end{center}
\end{table}

\section{INVESTIGATION OF THE PERIOD VARIATION}

The CCD image enables very high precision for measuring the mid-eclipse times (the errors are less than 17 seconds in our observations), and it is a very powerful tool for detecting the orbital variations of contact binaries. According to recent research, the period of a contact binary is not constant in many cases \citep{2017PASP..129l4202W,2018RAA....18...30Z,2018PASP..130d4201Z,2018NewA...62...20H,2018PASP..130g4201L}. The mass transfer between the two component stars, the presence of a tertiary object in orbit about the binary system as well as magnetic braking will cause period variations, which will eventually affect the evolution of the binary systems. To understand the interior/exterior dynamic interactions of PZ UMa, the O - C method was used, and all available times of light minima are listed in the first column of Table \ref{Minimum}. All light minima were observed with the CCD method.

\begin{table}[!h]
\caption{$O-C$ values of light minima for PZ UMa.}\label{Minimum}
\begin{center}
\small
\begin{tabular}{ccllrrcc}\hline\hline
  JD(Hel.)      &  p/s     &        Epoch        &     $(O-C)_1$      &   Error   & Ref.  \\
 (2400000+)     &          &                     &                    &           &        \\\hline
51337.7141	    &    p	   &         0           &        0           &           & 1      \\
55937.8090	    &    s	   &       17512.5       &     0.0165         &   0.0040  & 2      \\
55937.9390	    &    p	   &       17513.0	     &     0.0151         &   0.0006  & 2      \\
56017.6618	    &    s	   &       17816.5	   	 &     0.0164         &   0.0003  & 2      \\
56698.3779	    &    p	   &       20408.0	     &     0.0128         &   0.0013  & 3      \\
56706.3902	    &    s	   &       20438.5	     &     0.0136         &   0.0042  & 4      \\
56709.5398	    &    s	   &       20450.5	     &     0.0111         &   0.0015  & 4      \\
56711.3772	    &    s	   &       20457.5	     &     0.0098         &   0.0167  & 4      \\
57067.4213	    &    p	   &       21813.0	     &    -0.0008         &   0.0001  & 5      \\
57067.5546	    &    s	   &       21813.5	     &     0.0012         &   0.0001  & 5     \\
57067.6845	    &    p	   &       21814.0	     &    -0.0002         &   0.0002  & 5     \\
57498.0700	    &    s	   &       23452.5	     &    -0.0061         &   0.0002  & 6     \\
57498.2008	    &    p	   &       23453.0	     &    -0.0066         &   0.0001  & 6     \\
57499.1209	    &    s	   &       23456.5	     &    -0.0059         &   0.0001  & 6     \\
57733.2925	    &    p	   &       24348.0	     &    -0.0082         &   0.0001  & 6    \\
58452.3552	    &    s	   &       27085.5	     &    -0.0155         &   0.0002  & 6    \\\hline
\end{tabular}
\end{center}
\textbf
{\footnotesize Reference:} \footnotesize (1) \citet{2005OEJV...12....1N}; (2) \citet{2012IBVS.6029....1D}; (3) \citet{2014IBVS.6118....1H};
(4) \citet{2015IBVS.6149....1H}; (5) \citet{2017OEJV..179....1J}; (6) The present work.
\end{table}

\begin{figure}[!ht]
\begin{center}
\includegraphics[width=12cm]{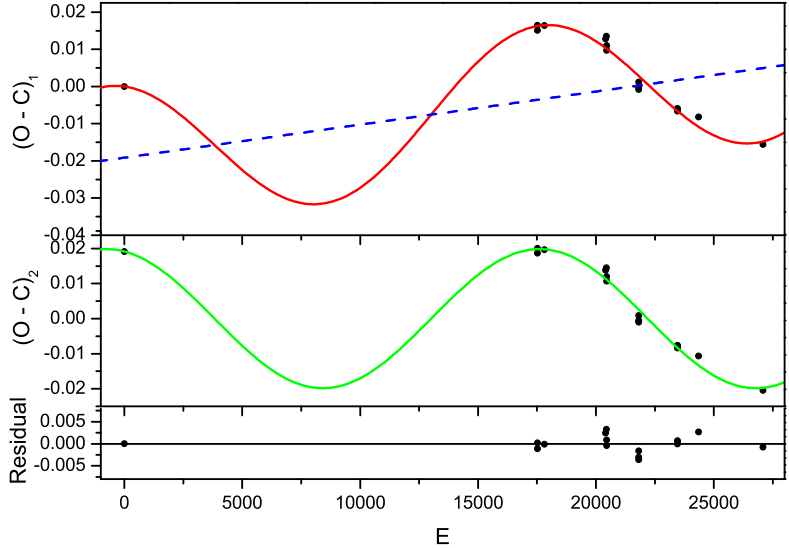}
\caption{The O - C curve of PZ UMa.}\label{O-C}
\end{center}
\end{figure}

The epoch (HJD = 2451337.7141) published by \citet{2005OEJV...12....1N} was used as the initial epoch, and period (P = 0.262674 days) in the O - C gateway\footnote{http://astro.sci.muni.cz/variables/ocgate/} was adopted. The corresponding epoch and $(O-C)_1$ values are listed in the third and fourth columns, which are also displayed in the upper panel of Fig. \ref{O-C}. To fit the observed light minima, a corrected linear ephemeris superimposed with a sine-like part was applied. The new ephemeris is:
\begin{equation}\label{New_ephemeris}
\begin{array}{lll}
Min. I=2451337.6950(\pm0.0004)+0.26267489(\pm0.00000002)\times{E}
        \\0.0198(\pm0.0002)\sin[0.0196^{\circ}(\pm0.0002)\times{E}+105.0^{\circ}(\pm0.6)]
\end{array}
\end{equation}

As shown in Fig. \ref{O-C}, the dots in the upper panel are the O - C values based on the initial ephemeris (Min. I = 2451337.7141 + 0.262674$\times$E), and the red line shows the calculated ones based on Equation \ref{New_ephemeris}. The dashed line represents corrections to the initial linear ephemeris. The sine-like curve seems more intuitive when the corrections are removed, which are shown in the middle part of Fig. \ref{O-C}. The fitted residuals are shown in the lowest part. The sine-like part in the O - C curve reveals that the period of PZ UMa contains a periodic variation, with a period of 13.22 years and an amplitude of 0.0198 days.

\section{MODELING OF LIGHT CURVES}

\begin{figure}[!ht]
\begin{center}
\includegraphics[width=12cm]{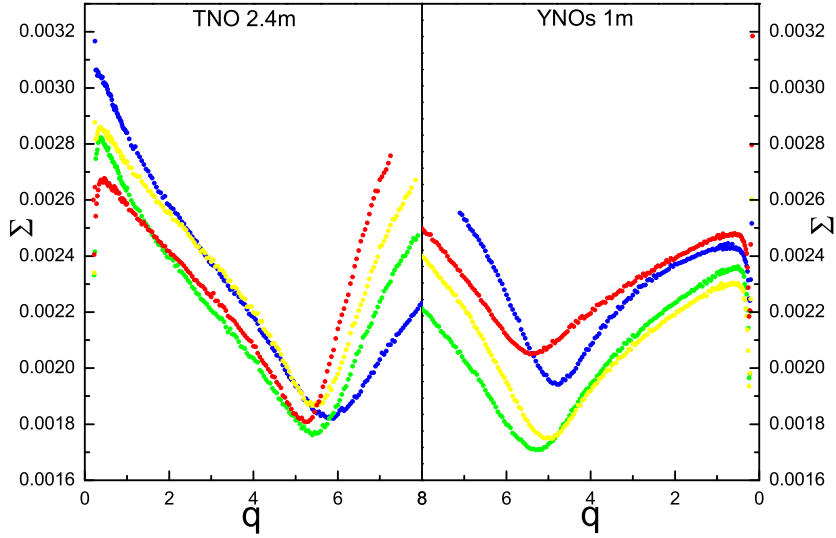}
\caption{The q-search results for light curves obtained with the TNO 2.4m (left) and YNOs 1m (right). The blue, green, yellow and red color refer to the $B$, $V$, $R_c$ and $I_c$ filters respectively.}\label{q-search}
\end{center}
\end{figure}

\begin{figure}[!ht]
\begin{center}
\includegraphics[width=12cm]{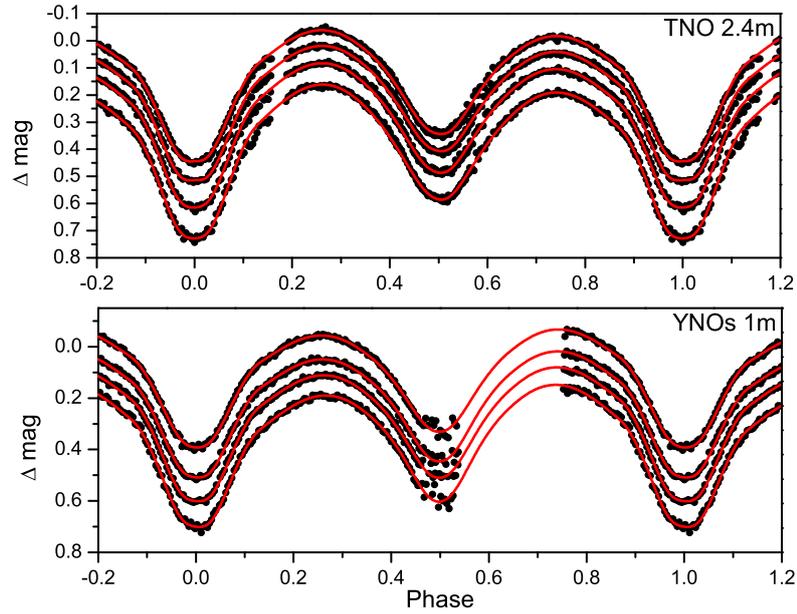}
\caption{The black solid circles are observational light curves obtained using the TNO 2.4m and YNOs 1m. The red lines are theoretical light curves calculated with the Wilson-Devinney program.}\label{LC_Cal}
\end{center}
\end{figure}

\begin{figure}[!ht]
\begin{center}
\includegraphics[width=13cm]{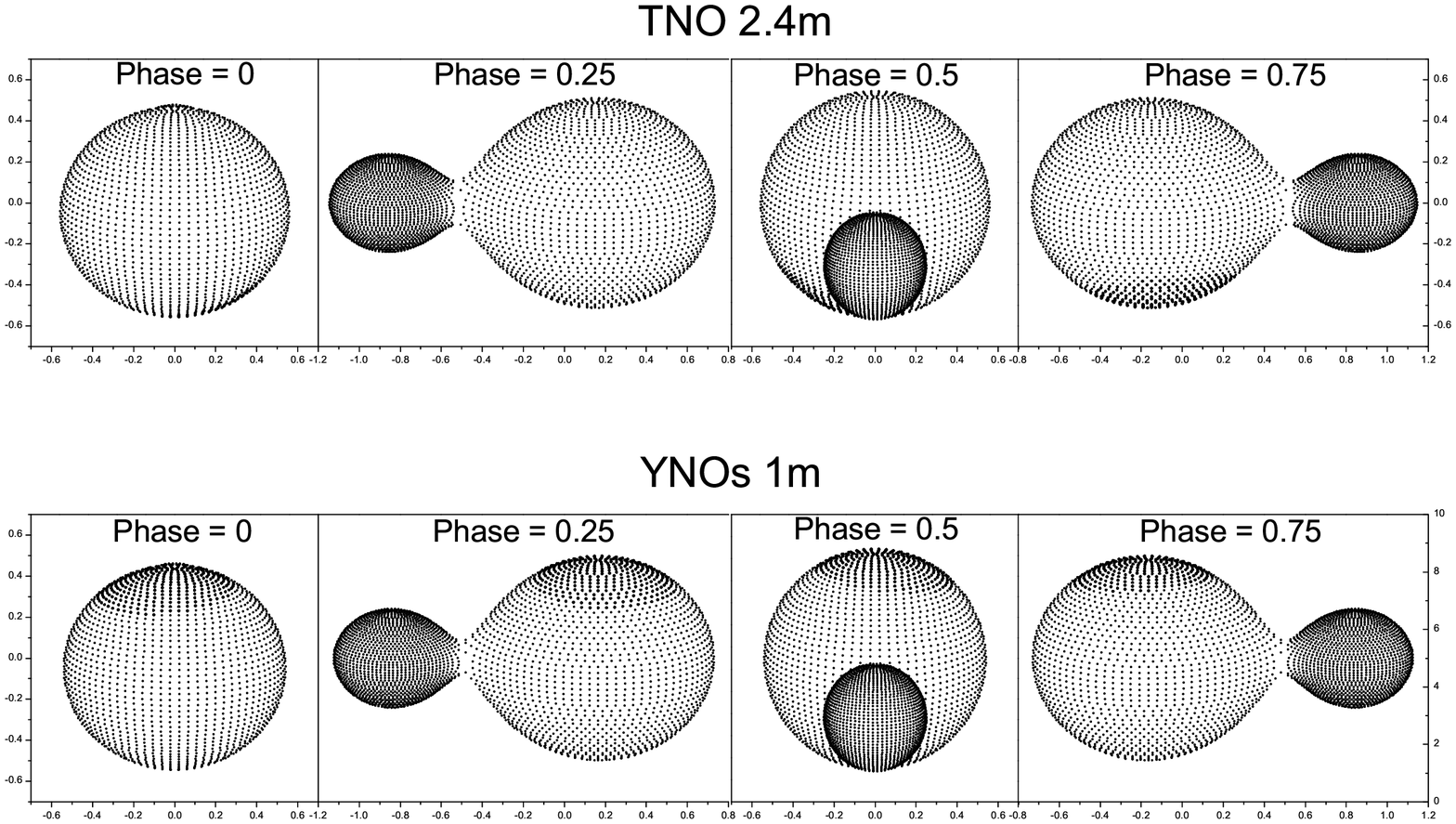}
\caption{The geometrical configurations at Phase = 0, Phase = 0.25, Phase = 0.5 and Phase = 0.75.}\label{CF}
\end{center}
\end{figure}

To determine the physical parameters of PZ UMa, the two sets of light curves derived from observations made with the TNO 2.4m and YNOs 1m were modeled with the Wilson-Devinney program \citep{2007ApJ...661.1129V,2014ApJ...780..151W}. The effective temperature of the primary star ( when eclipsed at primary minima ) was fixed at $T_{1} = 5430K$. The well known q-search method was used to search for a proper initial mass ratio ($q = M_{2}/M_{1}$). The q-search diagram is displayed in Fig. \ref{q-search}. It is obvious that PZ UMa is a W-subtype contact binary system. Then, the mass ratio is set to be a free parameter, the same as the orbital inclination ($i$), the dimensionless surface potential of the two stars ($\Omega_{1}=\Omega_{2}$ for contact mode ), the effective temperature of the secondary star ($T_{2}$), the monochromatic luminosity of the primary star ($L_{1}$), and the latitude ($\theta$) and longitude ($\psi$) of spots. As third light ($l_{3}$) has been detected in many close binaries \citep{2008AJ....136.2493Q,2013AJ....146...38Q,2015PASJ...67...98Z,2016AdAst2016E...7Z}, we try to set it as a free parameter. However, we can not get a converged solution with a positive parameter value of third light while running the Wilson-Devinney program, which may imply that there is no third light in PZ UMa. Firstly, we run the light curves in different bands separately, and the solutions are listed from column 2 to column 5 in Table \ref{WD_results1} and Table \ref{WD_results2}. Then, the light curves are fitted simultaneously to get the final solutions of PZ UMa, which are listed in the last column of Table \ref{WD_results1} and Table \ref{WD_results2}. The errors for each parameter listed in the last column are calculated from the solutions that we run the light curves separately. The theoretical light curves are plotted in Fig. \ref{LC_Cal}. The geometrical configurations at Phase = 0, Phase = 0.25, Phase = 0.5 and Phase = 0.75 are plotted in Fig. \ref{CF}. The positions of spots are also marked. It is evident that PZ UMa is a totally eclipsed binary system, and the smaller star is totally eclipsed at the primary minimum (Phase = 0).

\begin{table}[!h]
\begin{center}
\caption{Photometric solutions based on observations with the TNO 2.4m telescope.}\label{WD_results1}
\footnotesize
\begin{tabular}{lllllllll}
\hline\hline
Parameters              &    $B$                      &    $V$                      &    $R_c$                    &   $I_c$              &   $BVR_cI_c$ \\\hline
$T_{1}(K)   $           & 5430(fixed)                 & 5430(fixed)                 & 5430(fixed)                 & 5430(fixed)          & 5430(fixed)\\
q ($M_2/M_1$ )          & 5.65($\pm0.08$)             & 5.40($\pm0.09$)             & 5.59($\pm0.08$)             & 5.29($\pm0.06$)      & 5.62($\pm0.16$)\\
$i(^{\circ})$           & 74.0($\pm0.3$)              & 74.6($\pm0.3$)              & 75.1($\pm0.3$)              & 75.3($\pm0.2$)       & 74.8($\pm0.6$)\\
$\Omega_{1}=\Omega_{2}$ & 9.69($\pm0.10$)             & 9.41($\pm0.10$)             & 9.64($\pm0.09$)             & 9.27($\pm0.07$)      & 9.67($\pm0.18$)\\
$T_{2}(K)$              & 4984($\pm7$)                & 4951($\pm8$)                & 4903($\pm10$)               & 4889($\pm10$)        & 4972($\pm18$)\\
$\Delta T(K)$           & 446                         & 479                         & 527                         & 541                  & 458 \\
$T_{2}/T_{1}$           & 0.918($\pm0.001$)           & 0.912($\pm0.001$)           & 0.903($\pm0.002$)           & 0.900($\pm0.002$)    & 0.916($\pm0.003$)\\
$L_{1}/(L_{1}+L_{2}$)   & 0.288($\pm0.002$)           & 0.277($\pm0.002$)           & 0.262($\pm0.001$)           & 0.260($\pm0.001$)    &                  \\
$L_{1}/(L_{1}+L_{2}$) ($B$)  &                        &                             &                             &                      & 0.292($\pm0.001$)\\
$L_{1}/(L_{1}+L_{2}$) ($V$)  &                        &                             &                             &                      & 0.265($\pm0.001$)\\
$L_{1}/(L_{1}+L_{2}$) ($R_c$)&                        &                             &                             &                      & 0.248($\pm0.001$)\\
$L_{1}/(L_{1}+L_{2}$) ($I_c$)&                        &                             &                             &                      & 0.237($\pm0.001$)\\
$r_{1}(pole)$           & 0.238($\pm0.001$)           & 0.240($\pm0.001$)           & 0.238($\pm0.001$)           & 0.242($\pm0.001$)    & 0.238($\pm0.002$)\\
$r_{1}(side)$           & 0.250($\pm0.001$)           & 0.252($\pm0.001$)           & 0.249($\pm0.001$)           & 0.253($\pm0.001$)    & 0.249($\pm0.002$)\\
$r_{1}(back)$           & 0.297($\pm0.002$)           & 0.298($\pm0.002$)           & 0.296($\pm0.002$)           & 0.299($\pm0.002$)    & 0.295($\pm0.004$)\\
$r_{2}(pole)$           & 0.506($\pm0.005$)           & 0.504($\pm0.005$)           & 0.507($\pm0.004$)           & 0.502($\pm0.004$)    & 0.507($\pm0.009$)\\
$r_{2}(side)$           & 0.556($\pm0.008$)           & 0.554($\pm0.008$)           & 0.558($\pm0.007$)           & 0.551($\pm0.006$)    & 0.559($\pm0.015$) \\
$r_{2}(back)$           & 0.581($\pm0.010$)           & 0.580($\pm0.011$)           & 0.584($\pm0.009$)           & 0.577($\pm0.007$)    & 0.584($\pm0.019$)\\
$f$                     & $40.7\,\%$($\pm$15.5\,\%$$) & $38.4\,\%$($\pm$15.9\,\%$$) & $39.4\,\%$($\pm$13.4\,\%$$) & $37.8\,\%$($\pm$11.0\,\%$$) & $38.8\,\%$($\pm$28.2\,\%$$)\\
$\theta(^{\circ})$      & 157.7($\pm1.0$)             & 162.6($\pm1.1$)             & 162.6($\pm1.0$)             & 162.6($\pm1.0$)      & 161.3($\pm2.1$)\\
$\psi(^{\circ})$        & 288.2($\pm3.5$)             & 287.0($\pm5.3$)             & 283.6($\pm5.1$)             & 280.1($\pm4.5$)      & 281.9($\pm9.3$)\\
$r$(rad)                & 0.69(fixed)                 & 0.69(fixed)                 & 0.69(fixed)                 & 0.69(fixed)          & 0.69(fixed)\\
$T_f$                   & 0.81(fixed)                 & 0.81(fixed)                 & 0.81(fixed)                 & 0.81(fixed)          & 0.81(fixed)\\
$\Sigma{\omega(O-C)^2}$ & 0.00168                     & 0.00179                     & 0.00172                     & 0.00167              & 0.00355\\
\hline
\hline
\end{tabular}
\end{center}
\end{table}

\begin{table}[!h]
\begin{center}
\caption{Photometric solutions based on observations with the YNOs 1m telescope.}\label{WD_results2}
\footnotesize
\begin{tabular}{lllllllll}
\hline\hline
Parameters              &    $B$                      &    $V$                      &    $R_c$                    &   $I_c$              &   $BVR_cI_c$ \\\hline
$T_{1}(K)   $           & 5430(fixed)                 & 5430(fixed)                 & 5430(fixed)                 & 5430(fixed)          & 5430(fixed)\\
q ($M_2/M_1$ )          & 5.06($\pm0.17$)             & 5.29($\pm0.09$)             & 5.04($\pm0.08$)             & 5.40($\pm0.12$)      & 5.13($\pm0.24$)\\
$i(^{\circ})$           & 74.9($\pm0.4$)              & 74.7($\pm0.3$)              & 75.1($\pm0.3$)              & 74.1($\pm0.4$)       & 74.7($\pm0.7$)\\
$\Omega_{1}=\Omega_{2}$ & 9.10($\pm0.19$)             & 9.35($\pm0.10$)             & 9.05($\pm0.10$)             & 9.47($\pm0.14$)      & 9.17($\pm0.28$)\\
$T_{2}(K)$              & 5199($\pm29$)               & 5081($\pm16$)               & 5093($\pm19$)               & 5000($\pm30$)        & 5137($\pm49$)\\
$\Delta T(K)$           & 231                         & 349                         & 337                         & 430                  & 293 \\
$T_{2}/T_{1}$           & 0.957($\pm0.005$)           & 0.936($\pm0.003$)           & 0.938($\pm0.003$)           & 0.921($\pm0.006$)    & 0.946($\pm0.009$)\\
$L_{1}/(L_{1}+L_{2}$)   & 0.242($\pm0.003$)           & 0.247($\pm0.002$)           & 0.241($\pm0.002$)           & 0.237($\pm0.003$)    &                  \\
$L_{1}/(L_{1}+L_{2}$) ($B$)  &                        &                             &                             &                      & 0.256($\pm0.001$)\\
$L_{1}/(L_{1}+L_{2}$) ($V$)  &                        &                             &                             &                      & 0.240($\pm0.001$)\\
$L_{1}/(L_{1}+L_{2}$) ($R_c$)&                        &                             &                             &                      & 0.230($\pm0.001$)\\
$L_{1}/(L_{1}+L_{2}$) ($I_c$)&                        &                             &                             &                      & 0.224($\pm0.001$)\\
$r_{1}(pole)$           & 0.240($\pm0.002$)           & 0.238($\pm0.001$)           & 0.241($\pm0.001$)           & 0.237($\pm0.002$)    & 0.239($\pm0.003$)\\
$r_{1}(side)$           & 0.251($\pm0.002$)           & 0.249($\pm0.001$)           & 0.252($\pm0.001$)           & 0.248($\pm0.002$)    & 0.250($\pm0.003$)\\
$r_{1}(back)$           & 0.290($\pm0.002$)           & 0.290($\pm0.002$)           & 0.293($\pm0.002$)           & 0.290($\pm0.002$)    & 0.290($\pm0.004$)\\
$r_{2}(pole)$           & 0.494($\pm0.011$)           & 0.500($\pm0.005$)           & 0.495($\pm0.005$)           & 0.500($\pm0.008$)    & 0.497($\pm0.015$)\\
$r_{2}(side)$           & 0.539($\pm0.016$)           & 0.547($\pm0.008$)           & 0.540($\pm0.008$)           & 0.547($\pm0.011$)    & 0.544($\pm0.022$) \\
$r_{2}(back)$           & 0.564($\pm0.020$)           & 0.572($\pm0.010$)           & 0.565($\pm0.010$)           & 0.571($\pm0.014$)    & 0.569($\pm0.028$)\\
$f$                     & $23.0\,\%$($\pm$30.4\,\%$$) & $27.1\,\%$($\pm$15.8\,\%$$) & $26.3\,\%$($\pm$15.0\,\%$$) & $29.0\,\%$($\pm$22.6\,\%$$) & $24.2\,\%$($\pm$43.7\,\%$$)\\
$\theta(^{\circ})$      & 7.5($\pm0.7$)               & 5.2($\pm0.3$)               & 5.9($\pm0.3$)               & 5.5($\pm0.5$)        & 5.4($\pm1.0$)\\
$\psi(^{\circ})$        & 122.5($\pm6.7$)             & 99.9($\pm5.6$)              & 87.6($\pm6.0$)              & 64.9($\pm9.2$)       & 105.5($\pm14.0$)\\
$r$(rad)                & 0.69(fixed)                 & 0.69(fixed)                 & 0.69(fixed)                 & 0.69(fixed)          & 0.69(fixed)\\
$T_f$                   & 0.81(fixed)                 & 0.81(fixed)                 & 0.81(fixed)                 & 0.81(fixed)          & 0.81(fixed)\\
$\Sigma{\omega(O-C)^2}$ & 0.00205                     & 0.00167                     & 0.00173                     & 0.00221              & 0.00406\\
\hline
\hline
\end{tabular}
\end{center}
\end{table}

\section{DISCUSSION AND CONCLUSION}

The photometric solutions show that PZ UMa is a W-subtype contact binary with an extreme mass ratio ($1/q = 0.18$). Its totally eclipsed nature means that the photometric solutions are very reliable \citep{2017PASJ...69...37Z,2018RAA....18...59Z}. Normally, W UMa type contact binaries share a surface and the effective temperatures of the component stars are almost equal. As for PZ UMa, the temperature difference between the primary star and the secondary one is as large as several hundreds Kelvin. However, its fill out factor ($f$) is about $40\,\%$. This may indicate that energy transfer rate in PZ UMa is less efficient than other contact binaries and PZ UMa may have some similarities with the B-type systems or poor thermal contact binaries \citep{2004A&A...426.1001C}. PZ UMa may evolve into a extreme mass ratio overcontact binary system \citep{2008AJ....136.1940Q,2017PASJ...69...79L,2018AJ....156..199S}, and merge eventually as a new luminous red nova phenomenon \citep{1994IAUC.5942....1H,2003ApJ...582L.105S,2011A&A...528A.114T,2013A&A...555A..16T,2016RAA....16...68Z}.

The effective temperature of the secondary is $4972(18)K$, which corresponds to a K1 spectral type Main Sequence star. Therefore, the mass of the secondary star is estimated to be $M_{2} = 0.77(2)M_\odot$ \citep{Cox2000}. The mass of the primary star is calculated to be $M_{1} = 0.14(1)M_\odot$ while the photometric mass ratio $q = 5.62(16)$ is considered. Based on Kepler's third law, the orbital semi-major axis is determined to be a = $1.67(2)R_\odot$. The radii and luminosities of the two stars are listed as follows: $R_{2} = 0.92(1)R_\odot$, $R_{1} = 0.43(1)R_\odot$, $L_{2} = 0.46(2)L_\odot$ and $L_{1} = 0.15(3)L_\odot$. However, radial velocity curves are needed to determine the absolute parameters with much higher credibility.

The O - C curve reveals a periodic variation with a period of 13.22 years and an amplitude of 0.0198 days, which may be explained by the light travel time effect (LTTE) caused by a tertiary component. If we assume that the tertiary component was coplanar with the orbital plane of PZ UMa. Then, the distance from the binary to the common mass center of the triple system was $a_{12} = 3.43 AU$, and the mass function was $f(m) = 0.23 M_\odot$. The mass of the tertiary component was calculated to be $M_{3} = 0.88 M_\odot$, and its orbital radius was $a_{3} = 3.67 AU$. It is reported that many close binaries are actually triple systems \citep{2007AJ....134.1769Q,2011AJ....142..124Z,2014AJ....147...98L}, or even quadruple systems \citep{2013AJ....145...39Z,2015AJ....149..120L,2018PASP..130h4205Z}. What's more, extraneous eclipses are observed in some eclipsing multiple stars \citep{2017RAA....17...22Z,2018A&A...610A..72Z}. \citet{2014ApJ...793..137N} claimed that the ``eccentric Kozai-Lidov" mechanism would result in a tight inner binary in a multiple star system. The more interesting thing is that the minimum mass of the tertiary is calculated to be $M_{3min} = 0.84 M_\odot$ with the largest orbital radius to be $a_{3max} = 3.72 AU$. This means that the mass of the tertiary component is even larger than the two component stars of PZ UMa. However, we have not detected third light ($l_{3}$) through photometric analysis, which implies that the tertiary may be a compact object such as a white dwarf or a neutron star. \citet{2017RAA....17...87Q} \& \citet{2018ApJS..235....5Q} proposed that a compact tertiary component may even contaminate the inner binary during the early evolutionary stage of a multiple star system, which can explain the formation of a contact binary with a very high metallicity.

The O'Connell effect is commonly reported in W UMa type binaries \citep{2016NewA...47....3Z,2016PASJ...68..102X,2018PASJ...70...87Z}. It is widely accepted that the formation of the O'Connell effect is closely related to stellar magnetic activities and circumbinary dust \citep{2003ChJAA...3..142L}. Research on the O'Connell effect will help us to understand the magnetic activities of binary stars and the characteristics of circumbinary dust. Both of our two sets of light curves show the O'Connell effect. The light variations (Max I - Max II) between the two maxima are recorded and listed in Table \ref{MaxI-MaxII}, where Max I refers to the light maximum at phase 0.25 following the primary light minimum and Max II is the maximum at phase 0.75. The O'Connell effect observed by the TNO 2.4m was a positive one. However, it changed to be a negative one when it was observed with the YNOs 1m just ten months later. It is quite striking that the O'Connell effect changes over such a short time-scale. Considering that both of the components are late-type stars, we believe that magnetic activity is the most possible explanation in this case. Spot modelling was used to fit the observed light curves, and the positions of spots are marked in Fig. \ref{CF}. It is easy to see that the migration of the spot from the ``South pole" to ``North pole" accounts for the changes in the O'Connell effect. The O'Connell effect appears with late-type contact binaries quite frequently. Late-type contact binaries with the O'Connell effect are ideal targets for the study of stellar magnetic activities. They may help us to understand the formation, migration and disappearance of spots.

\begin{table}[!ht]\small
\begin{center}
\caption{Light variations of the two maxima (Max I - Max II).}\label{MaxI-MaxII}
\begin{tabular}{cccccc}\hline\hline
  Filter   &    TNO 2.4m     &  YNOs 1m   \\\hline\hline
   $B$     &     -0.028      &   0.042     \\
   $V$     &     -0.021      &   0.031     \\
  $R_C$    &     -0.022      &   0.031     \\
  $I_C$    &     -0.021      &   0.024     \\
\hline\hline
\end{tabular}
\end{center}
\end{table}

\bigskip

\vskip 0.3in \noindent
We thank Mr. Wayne Orchiston for improving the manuscript. This research was supported by the Chinese Natural Science Foundation (Grant No. 11703080 and 11703082) and the Yunnan Natural Science Foundation (No. 2018FB006). It was part of the research activities at the National Astronomical Research Institute of Thailand (Public Organization). This work has made use of data from the European Space Agency (ESA) mission {\it Gaia} ({https://www.cosmos.esa.int/gaia}), processed by the {\it Gaia} Data Processing and Analysis Consortium (DPAC, {https://www.cosmos.esa.int/web/gaia/dpac/consortium}). Funding for the DPAC has been provided by national institutions, in particular the institutions participating in the {\it Gaia} Multilateral Agreement. This research has made use of ``Aladin Sky Atlas" developed at CDS, Strasbourg Observatory, France.


\end{document}